\documentclass[useAMS]{nature}
\usepackage{hyperref} 
\usepackage[sort,numbers]{natbib}
\usepackage{xcolor}
\usepackage[normalem]{ulem}
\usepackage{amsmath}
\usepackage{mathtools}  
\usepackage{graphicx}
\makeatletter
\let\saved@includegraphics\includegraphics
\AtBeginDocument{\let\includegraphics\saved@includegraphics}

\makeatother

\hypersetup{
    colorlinks=true,
    linkcolor=blue,
    filecolor=magenta,      
    urlcolor=cyan,
    citecolor=blue
}
\newcommand{\mnras}{MNRAS}
\newcommand{\apj}{ApJ}
\newcommand{\araa}{ARA\&A}
\newcommand{\aap}{A\&A}
\newcommand{\aj}{AJ}
\newcommand{\apjl}{ApJL}
\newcommand{\apjs}{ApJS}
\newcommand{\pasp}{PASP}
\newcommand{\ao}{ApOpt}

\newcommand{\ha}{\hbox{H$\alpha$}}
\newcommand{\hb}{\hbox{H$\beta$}}
\newcommand{\oii}{\hbox{[O\,{\sc ii}]}}
\newcommand{\oiii}{\hbox{[O\,{\sc iii}]}}

\newcommand{\nii}{\hbox{[N\,{\sc ii}]}}

\title{Unusual integrated metallicity profile of our Milky Way}
\author{Jianhui Lian$^{1}$, Maria Bergemann$^{1}$, Annalisa Pillepich$^{1}$, Gail Zasowski$^{2}$, Richard R. Lane$^{3}$\\
\small $^{1}${Max Planck Institute for Astronomy, 69117, Heidelberg, Germany}\\
\small $^{2}${Department of Physics \& Astronomy, University of Utah, Salt Lake City, UT 84112, USA}\\
\small $^{3}${Instituto de Astrofısica, Pontificia Universidad Cat\'olica de Chile, Av. Vicuna Mackenna 4860, 782-0436 Macul, Santiago, Chile}\\
}
\begin{document}
\maketitle



{\bf The heavy element abundance profiles in galaxies place stringent constraint on galaxy growth and assembly history. Low-redshift galaxies generally have a negative metallicity gradient in their gas \citep{sanchez2014-gas-gradient,belfiore2017} and stars \citep{sanchez-Blazquez2014,goddard2017b}. Such gradients are thought to be a natural manifestation of galaxy inside-out formation \citep{Boissier1999,sanchez2014-gas-gradient}. 
As the Milky Way is currently the only spiral galaxy in which we can measure temporally-resolved chemical abundances, it enables unique insights into the origin of metallicity gradients and their correlation with the growth history of galaxies. However, until now, these unique abundance profiles had not been translated into the integrated-light measurements needed to seamlessly compare with the general galaxy population. Here we report the first measurement of the light-weighted, integrated stellar metallicity profile of our Galaxy. We find that the integrated metallicity profile of the Milky Way has a `$\wedge$'-shape broken metallicity profile, with a mildly positive gradient inside a Galactocentric radius of 7 kpc and a steep negative gradient outside. This metallicity profile appears unusual when compared to Milky Way-mass star-forming galaxies observed in the MaNGA survey and simulated in the TNG50 cosmological simulation. The analysis of the TNG50 simulated galaxies suggests that the Milky Way's positive inner gradient may be due to an inside-out quenching process. The steep negative gradient in the outer disc, however, is challenging to explain in the simulations. Our results suggest the Milky Way may not be a typical spiral galaxy for its mass regarding metallicity distribution and thus offers insight into the variety of galaxy enrichment processes.}  

%
%

Our home galaxy, the Milky Way, provides unique and strict constraints to galaxy formation and evolution because of the detailed, temporally-resolved observations that we can obtain from the inside. However, the integrated properties of the Milky Way are poorly understood and this limits a detailed comparative analysis of the properties of the Milky Way in the context of the general galaxy population, for the vast majority of which only integrated properties are measurable.

With the recent advent of massive spectroscopic surveys, which are mapping millions of stars across the Galaxy, direct measurements of {integrated stellar population properties (e.g., elemental abundances) of the Milky Way are becoming possible}. 
In this work, we present the first measurement of the radial integrated stellar metallicity profile of our Galaxy, carefully accounted for the selection function of the data, and {perform direct comparison with other similar-mass, star-forming galaxies both in the local Universe and in cosmological simulations of galaxy formation}.

We determine the integrated stellar metallicity\footnote{We use the standard astronomical notation to describe the metallicity of a star. For the stellar abundance of iron ([Fe/H]), ${\rm [Fe/H] = log(\frac{Fe/H}{Fe_{\odot}/H_{\odot}})}$}
profiles from 2 to 15~kpc of the Milky Way by using chemical abundances and ages of individual stars derived from spectra observed with APOGEE \citep{majewski2017} and astrometric data from Gaia \citep{gaia2016}. We transform the observations from a sample of targeted stars to the intrinsic, entire stellar population by correcting for the survey selection function for stars of different abundances separately. The obtained luminosity densities of intrinsic populations of different abundances are then used to calculate the light-weighted average stellar metallicity (Methods). 

When accounting for stars of all ages, the radial profile of the light-weighted stellar metallicity in the Milky Way shows a break at 6.9$\pm0.6$~kpc, with a positive slope of 0.031$\pm0.010$~dex/kpc within the break radius and a negative slope of $-0.052\pm$0.008~dex/kpc beyond it (Fig.1; Methods). This break, however, is not seen in the metallicity profiles of mono-age populations, which are either flat in the old age bin, or steep and negative in younger stellar populations. 
The fraction of total luminosity in the old (8-12~Gyr) and metal-poor stellar population decreases with radius, while the opposite is true for the younger, more metal-rich populations. This is consistent with the more radially compact morphology, i.e. a shorter scale length, of the old population \citep{bensby2011,bovy2012b}. This radially-varying contribution of the old, metal-poor vs. young, metal-rich stellar populations in the disc gives rise to the positive slope of metallicity profile in the inner Galaxy \citep{Schonrich2017}. For the same reason, the negative slope in the outer Galaxy reflects the gradient of the young and intermediate-age populations that dominate at larger radii.  

To compare the Milky Way with other galaxies, we measure the integrated stellar and gas-phase metallicity gradients of 544 face-on, star-forming galaxies with  Milky Way-like stellar mass ($|{\rm log(M_*/M_{MW}})|<0.2$~dex) in the MaNGA Integral Field Unit (IFU) survey \citep{bundy2015} (Methods). In addition, we compare our results with profiles of 134 Milky Way-mass, star-forming galaxies in the TNG50 cosmological hydrodynamical simulation \citep{Pillepich2019,Nelson2019}. {Figure 2 shows the total (i.e. integrated, left panel) and present-day (right panel)} metallicity profiles of these MaNGA and TNG50 galaxies in comparison to those of the Milky Way. 
{It is evident that the metallicity profiles of our Galaxy are inconsistent with the bulk of Milky Way-mass MaNGA and TNG50 galaxies, which -- when averaged across the populations -- generally show flatter radial metallicity distributions. The relatively flat stellar metallicity gradients of MaNGA galaxies are supported by independent measurements based on different stellar population synthesis codes \citep{boardman2020_MWA}, and are also consistent with those of nearby massive star-forming galaxies \citep{Saglia2018,Gregersen2015,Kudritzki2012,liu2022}, for which spectroscopic observations of resolved luminous stellar populations (e.g., red or blue supergiants) are available.} 
%

This difference in metallicity distribution between the Milky Way, on the one hand, and MaNGA and TNG50 galaxies, on the other, is statistically significant and holds also when galaxies are examined individually. We quantify the inner and outer gradients of the metallicity profiles of individual MaNGA and TNG50 galaxies with a broken linear function, the same as the Milky Way (Methods).
Fig. 3 compares the inner and outer metallicity gradients of the Milky Way with those of our MaNGA and TNG50 galaxies. A small fraction of galaxies in our MaNGA ($\sim$1$\%$) and TNG50 ($\sim$3\%) samples have a positive profile in the inner disc and a negative profile in the outer disc that are within 1$\sigma$ or steeper than the Milky Way. The majority of these galaxies are, however, characterised by a monotonic negative metallicity profile that tends to be flatter than that of the Milky Way, albeit with more frequent inner positive gradients and less diversity in TNG50 galaxies (possibly simply because of the limited sample size). 
Also the Milky Way's present-day metallicity profile, from the young stellar populations and with gradient of -0.049$\pm$0.005~dex consistent with observations of different individual types of young stellar objects \citep{braganca2019,minniti2020,zhang2021} and H~II regions \citep{wenger2019}, is significantly steeper than the typical gas-phase metallicity profiles of the MaNGA sample\citep{boardman2020_MWA} and than the present-day metallicity profiles of the TNG50 galaxies.

The analysis of the simulated galaxies suggests that there is an innate relationship between the shape of the galaxy metallicity profiles {and the disc structure}.
In the presence of broken metallicity profiles like is the case for our Milky Way, it is most intuitive to consider inner and outer discs of galaxies separately. To explore the origin of the metallicity gradients by marginalizing over the effect of galaxy size -- which can vary by up to a factor of 5 at fixed galaxy mass -- here we normalize the inner and outer metallicity gradients by the effective radius of each individual galaxy. Because of {an uncertainty} in the current measurement of Milky Way's size (Methods), here we adopt a range of effective radius of $3.4-6.7~$kpc \citep{bland2016}, corresponding to a scale length of $2-4~$kpc assuming a single exponential profile, and depict possible positions of the Milky Way in Fig. 4 assuming equally possible values within this range.

In the inner regions (Fig 4a), the Milky Way and TNG50 galaxies with Milky Way-like disc structure evolution 
show a positive metallicity gradient accompanied by a positive gradient of luminosity profile of the young population. This indicates reduced recent star formation in the innermost regions of these galaxies, and could be a manifestation of inside-out quenching of star formation \citep{wang2018,lin2019,hasselquist2019}, possibly due to AGN feedback as is the case for TNG50 galaxies \citep{Nelson2019, Nelson2021, Pillepich2021}. 

In the outer regions (Fig. 4b), however, TNG50 galaxies with Milky Way-like structure typically show a flatter metallicity gradient. 
The correlation between the evolution of structure and the stellar metallicity gradient in the outer disc cannot be confirmed for the Milky Way, because of the large uncertainty on the Milky Way's effective radius. If the currently estimated disc size is confirmed, the Milky Way - with its steep outer metallicity gradient - will likely represent an unusual galaxy for its mass and star formation activity that is not captured or realized by the TNG50 simulation. As suggested in empirical chemical evolution models or zoom-in simulations, such a steep outer metallicity gradient is possibly induced by the following processes under certain realization: inside-out ignition of star formation due to decreasing gas density with radius and inside-out disc growth due to increasing gas accretion timescale with radius \citep{chiappini2001,Schonrich2017}, radial gas inflow along the disc that carries enriched material inwards \citep{schonrich2009,chen2022}, and an abrupt, metal-free gas accretion event via e.g. a minor merger that dilutes the disc more at larger radii \citep{buck2020,agertz2021,lian2020b}. The metallicity gradient shaped by these processes depend on how they are implemented in the models. For instance, in the metal-free gas accretion scenario, the resulting dilution as a function of radius strongly depends on the radial distribution of gas infall and therefore the mass and orbit of the infalling satellite.  
These requirements could make the accretion-induced steep metallicity gradient uncommon. 

To summarise, we find that the integrated, light-weighted metallicity profile of the Milky Way is non-monotonic, with a positive gradient inside 7~kpc and a negative gradient outside. This metallicity profile is significantly different from the majority of Milky Way-mass star-forming galaxies both in the local Universe and state-of-the-art cosmological simulations. The overall shape of the gradient is set by the temporal evolution of the inner and outer disc components. The chemical structure of the inner galaxy can be explained by the inside-out quenching of star formation in the inner regions. This is consistent with observations of other Milky Way-like galaxies in the local volume and with properties of model galaxies in e.g. the TNG50 cosmological simulation. The structure of the outer regions of the Milky Way is, however, more challenging to explain in the accepted galaxy formation framework, as the steep metallicity outer profile, combined with the current estimate of the effective radius, makes our Galaxy an atypical system compared to a few hundred Milky Way-like galaxies realized in current cosmological simulations and observedations of Milky Way-like galaxies in the local Universe. This discrepancy may point either to the erroneous measurement of the Galactic disc size, which may be addressed with upcoming surveys like WEAVE, 4MOST, and SDSS-V, or to a particular physical process operating during the evolution of the Milky Way that is not captured or is not sufficiently frequent to be realized by the currently available simulated galaxies. 

\begin{figure*}
	\centering
	\includegraphics[width=12cm]{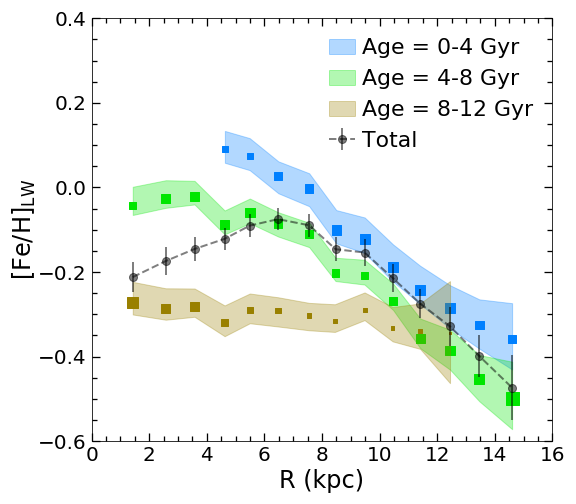}
	\caption{Average light-weighted stellar iron and magnesium abundance profiles of the Milky Way galaxy as a whole and of three different age populations. 
	The integrated metallicity of all ages and that in each age bin are average values of mono-abundance populations, weighted by their optical r-band luminosity (Methods). The size of the colourful squares indicates the fraction of the total luminosity at each radial bin contained in each mono-age component.        
	} 
	\label{z-prof-age}
\end{figure*} 

\begin{figure*}
	\centering
	\includegraphics[width=13cm]{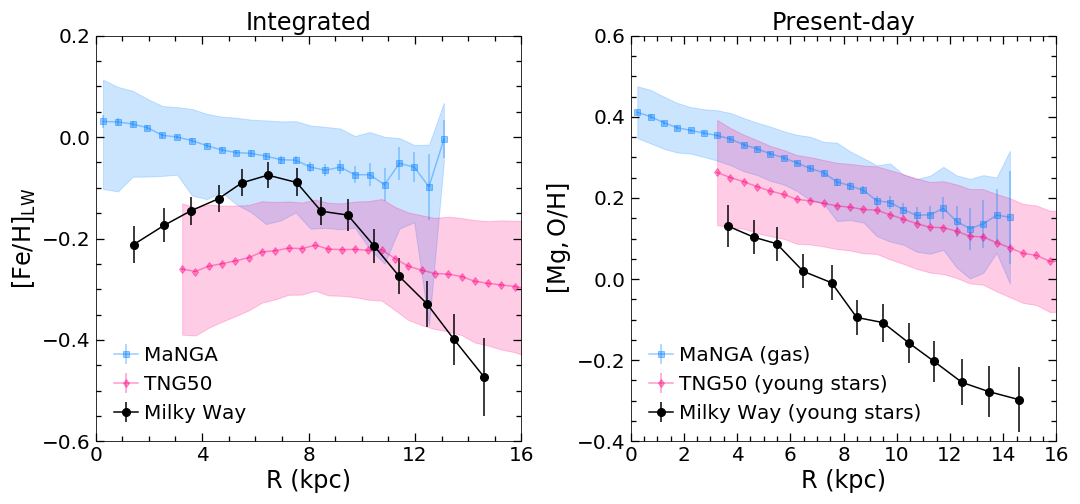}
	\caption{{\sl Left:} Average radial light-weighted stellar metallicity profile of the Milky Way (black) in comparison with those of  low-redshift Milky Way-mass star-forming galaxies in the MaNGA survey (blue) and in the TNG50 simulation (magenta), averaged across 544 and 134 galaxies, respectively. 
	Black errorbars indicate the stochastic uncertainty of the integrated stellar metallicity of the Milky Way.
	The blue and magenta shaded regions represent the 1~$\sigma$ scatter of the distributions of these Milky Way-mass star forming galaxies in the MaNGA survey and TNG50 simulation, respectively. The filled symbols and errorbars denote their median metallicity profile and the error of the median. {\sl Right:} Comparison among the same galaxy samples as on the left but now for their present-day metallicity gradients. For MaNGA galaxies, we use the oxygen abundance ([O/H]) measured in their ionized gas from optical emission lines (Methods) to represent their present-day metallicity. For the Milky Way and TNG50 galaxies, we use [Mg/H], which is tightly correlated with [O/H] \citep{ting2022}, of their young (0-4~Gyr) population to represent their present-day metallicity (Methods).   
	} 
	\label{z-prof}
\end{figure*} 

\begin{figure*}
	\centering
	\includegraphics[width=16cm]{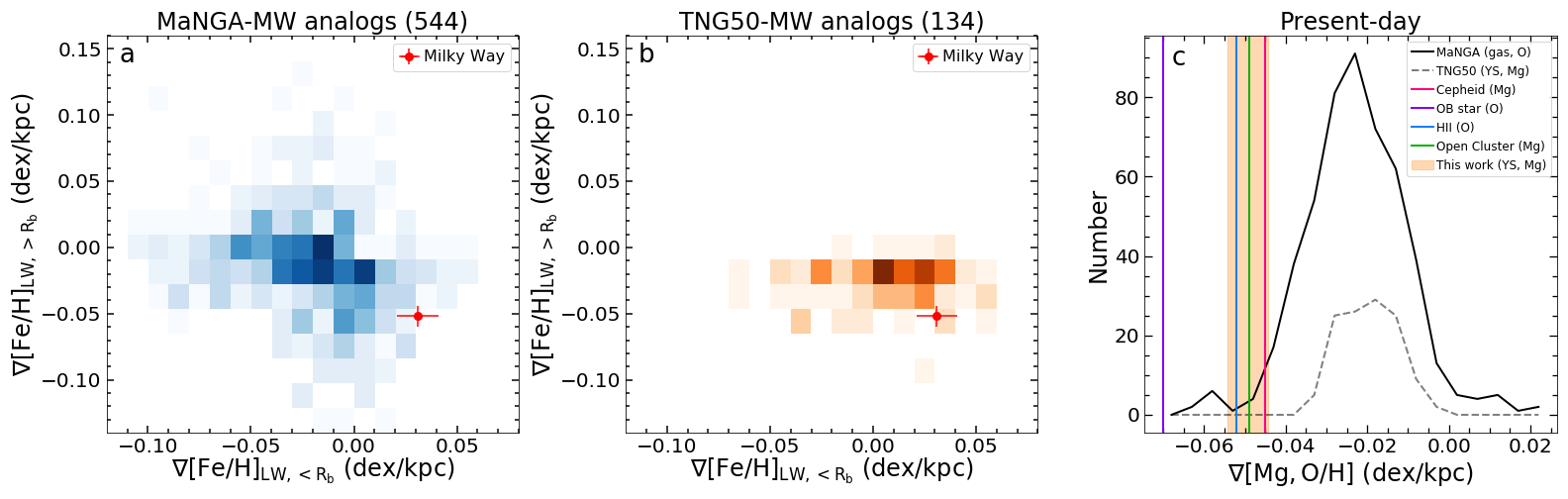}
	\caption{Comparison of the light-weighted average stellar metallicity gradients in- and outside the break radius ($\rm R_{b}$ -- panels a and b) and of the present-day gradient (panel c) of the Milky Way with those of Milky Way-mass star-forming galaxies in the MaNGA survey and the TNG50 simulation. In panels a and b, the distributions of MaNGA and TNG50 galaxies are colour coded by their number in each pixel. In both cases, the Milky Way is located at the bottom right edge of the distributions of the observed or simulated galaxy populations. In panel c, the shaded region indicates the 1~$\sigma$ stochastic uncertainty of the Milky Way's present-day gradient. For reference, we include the gradient of oxygen or magnesium abundance of different types of young objects in the Milky Way, including OB stars \citep{braganca2019}, Cepheids \citep{genovali2015}, open cluster \citep{magrini2017}, and H II regions \citep{wenger2019}. 
	} 
	\label{zgrad-hist2d}
\end{figure*} 

\begin{figure*}
	\centering
	\includegraphics[width=16cm]{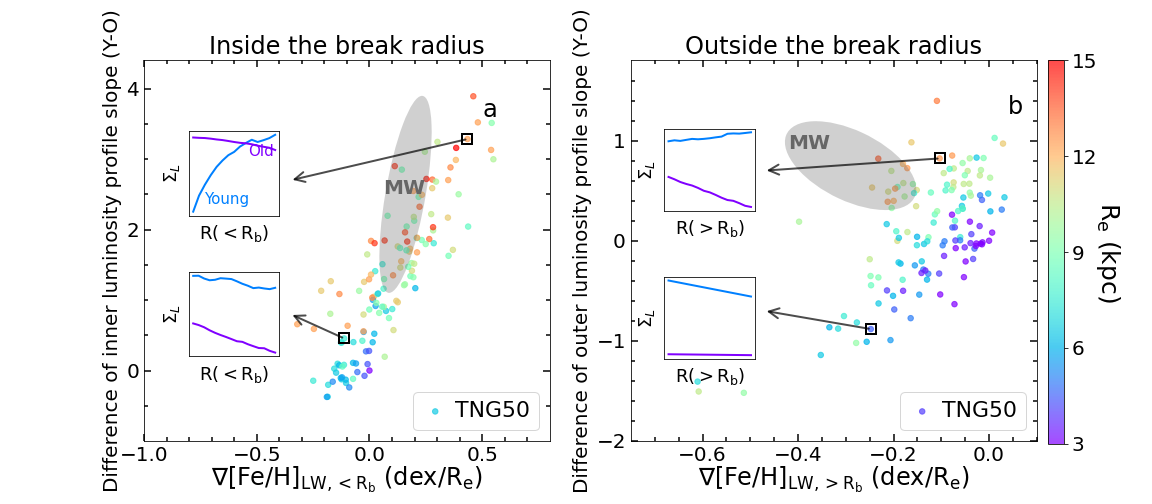}
	\caption{Correlation between the integrated stellar metallicity gradients {\sl normalized} to the galaxy effective radius and the disc structure evolution. The latter is quantified as the difference between the slope of radial luminosity surface density profiles ($\Delta {\rm log(\Sigma_L)/\Delta R}$) of young (0-4~Gyr) and old (8-12~Gyr) stellar populations, within- (panel a) and outside (panel b) the break radius identified in the metallicity radial profiles. Filled circles denote Milky Way-mass, star-forming galaxies in the TNG50 simulation colour-coded by their effective radius. Two insets on the left-hand side of each panel illustrate the luminosity surface density profiles of young and old populations in two example TNG50 galaxies (black squares): one with strong evolution at the top and one with nearly parallel growth (panel a) or inverse evolution (young population being more compact, panel b) at the bottom. The possible position of the Milky Way in these two diagrams is denoted by a shaded ellipse, which is stretched by the large uncertainty on the Milky Way's effective radius, in turn due to uncertainties in the disc scale length measurement and to the possible deviation from single exponential density profile at large distance. 
	} 
	\label{zgrad-structure}
\end{figure*} 

\newpage 

\section{Methods}

{\bf Data.} This work is based on the data from the last internal data release of the APOGEE survey after the SDSS-IV Data Release 16 (DR16)\citep{blanton2017,ahumada2020,jonsson2020}.  
APOGEE is a massive near-infrared, high-resolution spectroscopic survey \citep{majewski2017} that provides robust and precise stellar parameters and elemental abundances for more than a half million stars, mainly red giant branch stars, in nearly 1000 discrete fields that are semi-regularly distributed throughout the Galactic disc, bulge, and halo \citep{zasowski2013,zasowski2017,beaton2021,santana2021}. 
The observed sample is randomly selected, on a field-to-field basis, from candidates defined in the 2MASS $H$--$(J-Ks)_0$ colour-magnitude diagram. The stellar spectra are obtained using custom spectrographs \citep{wilson2019} with the 2.5 m Sloan Telescope
and the NMSU 1 m Telescope at the Apache Point Observatory \citep{gunn2006,holtzman2010}, and with the 2.5~m Ir\'en\'ee du~Pont telescope at Las Campanas Observatory \citep{bowen1973}. 
The spectra are reduced and chemical abundances (e.g., [Fe/H], [Mg/Fe]) and stellar parameters (e.g., log(g) and T$_{\rm eff}$) of individual stars are produced by custom pipelines using a new custom line list (ASPCAP)\citep{nidever2015,garcia2016,smith2021}.
The stellar ages and spectro-photometric distances are derived by applying the astroNN deep-learning code to the spectroscopic data from APOGEE and asterometric data from Gaia, and are provided in the astroNN Value Added Catalog \citep{mackereth2019,leung2019}.  

We have corrected the APOGEE abundances of Fe and Mg for the effects of non-local thermodynamic equilibrium (NLTE). This physical process is not captured by standard LTE models and is often taken into account explicitly \citep{Chaplin2020}. We used the Fe and Mg NLTE model atoms developed by Bergemann et al.\citep{Bergemann2012,Bergemann2017}. NLTE corrections were calculated for a grid of stellar model atmospheres covering the stellar parameter space of APOGEE observations for individual Mg I and Fe I lines, which are detectable in the H-band APOGEE spectra \citep{Souto2018}. The NLTE abundance corrections are typically within  $ +/- 0.10$~dex for Fe and Mg abundances in individual stars, but they amount to less than 0.02~dex for the integrated light-weighted abundances. The corrections for NLTE effect only change the metallicity gradient by less than 2\%.  

{\bf Integrated light-weighted metallicity measurements.} The integrated stellar metallicity in the Milky Way is derived using the density distribution of mono-abundance populations (MAPs) after carefully correcting for the APOGEE survey selection function \citep{lian2022b}. The process of correcting for the APOGEE selection function is described in greater detail in\citep{lian2022b}, but in summary:
Using PARSEC isochrones \citep{bressan2012} and the combined 3D extinction map \citep{bovy2016_dust}, we estimate the probability that a star, at a given Galactic position and [Fe/H] and [Mg/Fe] abundances,  would be selected as a candidate and then eventually  observed. The observed number density  of APOGEE stars at this position and abundance, divided by this observational probability, gives rise to the local density of the underlying population. This conversion from sampled to intrinsic number density is conducted for all individual MAPs at different Galaxy positions.  
We consider MAPs in the range of abundance  where the vast majority of the Milky Way's stars are located: [Fe/H] between -0.9 and +0.5, with bin width of 0.2~dex, and [Mg/Fe] between -0.1 and 0.4, with bin width of 0.1~dex. The luminosity density of each MAP is then obtained by sampling the PARSEC isochrones assuming a Kroupa IMF \citep{kroupa2001}. 
In total, we obtain 3056 individual luminosity measurements for MAPs spanning Galactocentric radii of 0-25~kpc and vertical distances of 0--14~kpc. 
 
With the luminosity density distribution, we first obtain the luminosity surface density of different MAPs as a function of radius by collapsing the density distribution in the vertical direction. We bin the density measurements of MAPs into a series of narrow radial bins, from 0 to 15~kpc with bin width of 1~kpc. For each radial bin, we fit the vertical density distribution with a single exponential profile and derive the surface luminosity density by integrating the best-fitted density model. These surface luminosity densities are used to weight each MAP to measure the average light-weighted metallicity of the Milky Way. These values then have the same physical meaning as  unresolved stellar metallicity measurements in external galaxies and therefore allows direct comparison between them.  
The average light-weighted metallicity is  calculated via:
\[
{\rm [Fe/H]_{LW} = \frac{\sum_i[Fe/H]_i*\sigma_{L,i}}{\sum_i \sigma_{L,i}}}, 
\]
where $\sigma_{\rm L,i}$ indicates the luminosity surface density of MAP $i$, and [Fe/H]$_i$ denotes the iron abundance of that MAP. The same calculation is performed to obtain the light-weighted Magnesium abundance ([Mg/H]), which is used in the comparison with the gas phase metallicity of galaxies that is usually represented by Oxygen abundance ([O/H]). The calculation is performed in the radial range 2-15~kpc where the vertical structure of all dominant MAPs are well determined.
The stochastic uncertainty of the integrated metallicity is estimated through Monte Carlo simulations, considering  uncertainties of abundances of each MAP and of the obtained surface mass density that is propagated from number density errors at each spatial position. As a conservative estimate, we assume the uncertainty of [Fe/H] and [Mg/Fe] in each MAP to be 0.1 and 0.05~dex, 
respectively, which are half of the bin width used for MAP definition.

Since the radial metallicity distribution in the Milky Way varies systematically with height from the disc plane\citep{hayden2015}, correcting the survey selection function is essential to derive the integrated average stellar metallicity and its radial distribution. However, the main result of this paper -- i.e., that the Milky Way presents a metallicity profile with a pronounced break -- is not significantly dependent on the correction for the selection function. The radial metallicity profiles obtained using APOGEE data without accounting for the selection function show similar behavior to those corrected for the selection function (Figure~5 vs. Figure~1), with a clear break at $r\sim6$~kpc, a positive gradient inside the break radius, and a strong negative gradient outside. Both the inner and outer gradients are steeper than the integrated stellar metallicity. This suggests that our results are robust against uncertainties in the correction for survey selection function, but such correction is necessary for an accurate understanding of the average metallicity distribution in the Milky Way.  

{\bf Metallicity profiles in age bins.} To investigate the origin of the broken integrated metallicity profile in the Galaxy, we study the metallicity profiles of stars at different ages. These time-resolved metallicity profiles are obtained by first unfolding the number density of each MAP at a given position along the age dimension using the observed age distribution. Then, for each age bin, we perform the same analysis above to derive the surface luminosities of MAPs and light-weighted average metallicity as a function of radius. Considering age uncertainties of $\sim$30\%, we consider three broad age bins from 0 to 12~Gyr with even steps of 4~Gyr.




{\bf Measurement of metallicity gradients.}
To measure the metallicity gradient of the Milky Way, because of the visibly non-monotonic profile, we fit a broken linear function, 
\[
y=
\begin{cases}
a_{\rm in}\times x + b ,x<{\rm R_b}\\
a_{\rm out}\times x + (a_{\rm in}-a_{\rm out}){\rm R_b}+b ,x>{\rm R_b},
\end{cases}
\]
to the profile with four free parameters: zero point ($b$), break radius (${\rm R_b}$), and gradient in- and outside ($a_{\rm in}$, $a_{\rm out}$) the break radius. The fit is performed in the radial range 2-15 kpc.  
The uncertainties of the best-fitted gradients are estimated through Monte Carlo simulations considering the stochastic uncertainty of integrated metallicity measurement at each radial bin. The best-fitted break radius is at 6.9$\pm0.6$~kpc. The effective radius of the Milky Way is rather uncertain, because of the significant uncertainty about the Milky Way's disc scale length \citep{bland2016} and the complexity of the disc density profile, which may deviate from a single exponential shape \citep{bovy2016,mackereth2017,yu2021,lian2022b}. This large uncertainty in the size measurement results in large uncertainty in the normalized metallity gradient of the Milky Way in dex/${\rm R_{e}}$. 


{\bf Comparison sample from the MaNGA survey.} We compare our findings of the metallicity gradients of the Milky Way with those of similar-mass, star-forming galaxies from the SDSS-IV MaNGA survey \citep{bundy2015} and the TNG50 cosmological simulation \citep{Pillepich2019,Nelson2019}. Raw MaNGA data are spectrophotometrically calibrated \citep{yan2016} and reducted by the Data Reduction Pipeline \citep{law2016}. From the final MaNGA Product Launch 11, we select 544 face-on galaxies (axis ratio $b/a > 0.5$, using $a$ and $b$ from the NSA catalog\citep{blanton2011}) with specific star formation rate of ${\rm sSFR > 10^{-11}~yr^{-1}}$ and $|\log{(M_*/M_{MW})}| < 0.2$~dex, assuming $\log{(M_{MW}/M_\odot)}$ = 10.76 \citep{licquia2016}. 
We take stellar metallicities in each galaxies from Firefly MaNGA value added catalog\citep{goddard2017b,parikh2018,neumann2021}, which uses the Firefly full spectrum fitting code \citep{wilkinson2017_ffly} and the MaStar stellar library \citep{maraston2020,yan2019}. The spaxels of each galaxy are binned using a Voronoi tessellation to ensure minimum spectra SNR of 10 \citep{cappellari2004}. We require each Voronoi-binned cell to have uncertainty of stellar metallicity measurement less than 0.5~dex. 


To make a fair comparison with the the Milky Way, we fit the same broken linear function to the metallicity profiles of these Milky Way-mass, star-forming MaNGA galaxies. The fits are performed over the radial range covered by MaNGA IFU bundle (0-1.5 effective radius for two thirds of MaNGA sample and 0-2.5 for the rest). The results 
are shown in Figures 3. Similarly flat stellar metallicity gradients of the {\it average} Milky Way-like MaNGA galaxies ($\sim0.02$~dex/kpc) are reported by \citet{boardman2020_MWA}, who utilized a different full spectrum fitting code (pPXF\citep{cappellari2004,cappellari2017}) and library (MILES \citep{sanchez-Blaquez2006}).
The consistent flat metallicity gradient on average based on the two sets of stellar metallicity measurements indicate that the results of this work are robust against potential systematics in extra-galactic stellar metallicity measurement from stellar population synthesis.  

We measure the gas metallicity of the same cells as above using emission line fluxes produced by the Data Analysis Pipeline \citep{belfiore2019,westfall2019} and two strong line metallicity calibrations, N2O2\citep{dopita2013} (\nii/\oii3727) and O3N2\citep{pettini2004} ((\oiii/\hb)/(\nii/\ha)), to account for systematics in gas metallicity measurement between different calibrations. For each cell, we require the SNR of the following strong emission lines to be above 5: (\oii$\lambda$3727,\hb,\oiii$\lambda\lambda$4959,5007,\ha,\nii$\lambda$6584), that their ratios place the cell in the star-forming region defined by the {conventional demarcation line\citep{kauffmann2003} in the} BPT diagram \citep{baldwin1981}. We correct for the galactic internal extinction using Balmer decrement method, assuming case B recombination with intrinsic \ha/\hb ratio of 2.87 \citep{osterbrock2006}, and 
a standard Milky Way extinction law \citep{cardelli1989}. For both gas metallicity calibrations, we obtain consistent metallicity gradients that are much flatter than the Milky Way.

{\bf Metallicity profiles of gas and young stars.} In the main text, we use the metallicity profile of young stars (0--4~Gyr) in the Milky Way and TNG50 galaxies to represent their present-day metallicity profile. To verify this approach we compare the metallicity profiles and gradients between the gas and young star particles in our TNG50 galaxies (Figure 6). The metallicity profile of young stars closely follows that of gas, although the latter is slightly steeper by 0.003~dex/kpc on average. This difference does not significantly affect the comparison in Fig.~3 in the main text. 

{\bf Comparison sample from the TNG50 simulation.} To compare the Milky Way results with those from current galaxy-formation models, we select Milky Way-like galaxies in the TNG50 simulation, by applying the same selection criteria (i.e. stellar mass and sSFR cuts) used for the MaNGA survey. TNG50 returns 134 
galaxies satisfying these criteria at the $z=0$ snapshot. 

TNG50 is a cosmological magnetohydrodynamical simulation for the formation and evolution of galaxies that encompasses a volume of about 50 comoving Mpc and that thus samples many thousands of galaxies across the mass spectrum, across galaxy types and environments \citep{Pillepich2019,Nelson2019}. In the simulation, processes like density-threshold star formation, stellar evolution, chemical enrichment, galactic winds generated by supernova explosions, gas cooling and heating, and seeding, growth and feedback from super massive black holes are all simultaneously followed \citep{Weinberger2017, Pillepich2018}, with an average mass and spatial resolution in the star-forming of galaxies of $8.5\times10^4\,M_{\odot}$ and 50-200 parsecs, respectively \citep{Pillepich2019, Pillepich2021}.
The TNG50 galaxies at $z=0$ have been shown to have structures and properties that are overall consistent with many observational findings: of relevance for the purposes of this comparison, these include the gas-phase metallicity gradients \citep{Hemler2021}, the radial star-formation-rate surface density profiles in comparison to MaNGA galaxies \citep{Motwani2020}, and the stellar sizes and overall stellar morphologies in comparison to SDSS data and others \citep{Zanisi2021}.
All this allows us to compare the case of the Milky Way to a relatively large set of simulated and reasonably-realistic galaxies. 

Following the same procedure as the Milky Way data, we obtain the integrated light-weighted metallicity profiles of the selected TNG50 galaxies and measure their metallicity gradients and break radii by fitting the light-weighted metallicity profiles of stellar particles within $\pm$4 kpc from the mid plane and in the 3-25 kpc range. As for the Galaxy we obtain profiles outwards of 2 kpc, we also apply a similar minimum radius for fitting the profiles of simulated galaxies. Because TNG50 galaxies are generally more extended than the Milky Way, we apply a maximum radius of 25~kpc to cover the simulated galaxies at least out to 2.5 effective radius. 
For their present-day metallicity gradient, to be consistent with the Milky Way, we use the metallicity gradient of young stars with ages of 0-4~Gyr. 
The results from these fits are shown in Figures 3 and 4. The effective radii of TNG50 galaxies are measured using the luminosity distribution at the $z=0$ snapshot.


%
%
%
\begin{figure*}
	\centering
	\includegraphics[width=12cm]{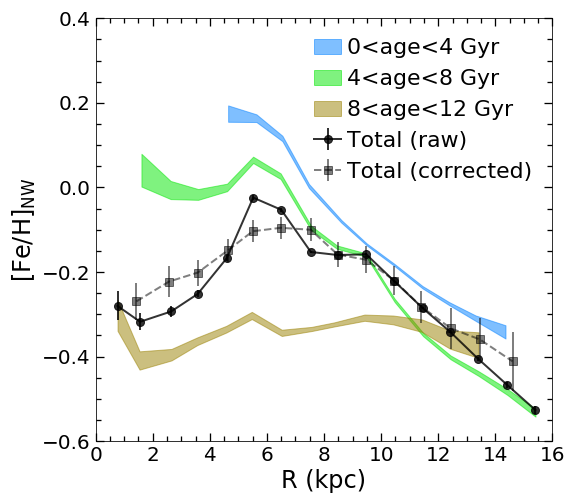}
	\caption{As in Fig.~1, i.e. radial average metallicity profile of total and mono-age populations of the Milky Way using APOGEE data, but without correcting for selection effects. The width of the shaded region indicate the error of the average metallicity. The integrated average metallicity profile that has been corrected for selection function (grey) is copied from Fig.~1 for reference. 
	} 
	\label{z-prof-age-raw}
\end{figure*} 

\begin{figure*}
	\centering
	\includegraphics[width=12cm]{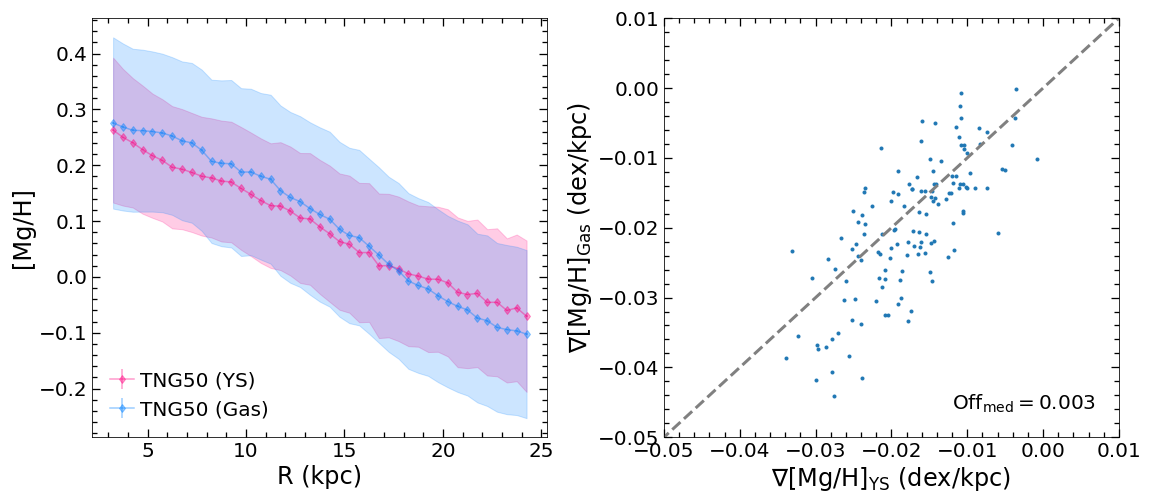}
	\caption{Comparison between the metallicity profiles (left) and gradients (right) of the gas and young (0--4~Gyr) star (YS) particles in our TNG50 galaxy sample. The metalliicty profile of young stars is similar to that of gas with the latter being slightly steeper by 0.003~dex/kpc on average. 
	} 
	\label{gas-ys}
\end{figure*} 





{\bf Acknowledgement}

Funding for the Sloan Digital Sky Survey IV has been provided by the Alfred P. Sloan Foundation, the U.S. Department of Energy Office of Science, and the Participating Institutions. SDSS-IV acknowledges
support and resources from the Center for High-Performance Computing at
the University of Utah. The SDSS web site is www.sdss.org.

SDSS-IV is managed by the Astrophysical Research Consortium for the 
Participating Institutions of the SDSS Collaboration including the 
Brazilian Participation Group, the Carnegie Institution for Science, 
Carnegie Mellon University, the Chilean Participation Group, the French Participation Group, Harvard-Smithsonian Center for Astrophysics, 
Instituto de Astrof\'isica de Canarias, The Johns Hopkins University, Kavli Institute for the Physics and Mathematics of the Universe (IPMU) / 
University of Tokyo, the Korean Participation Group, Lawrence Berkeley National Laboratory, 
Leibniz Institut f\"ur Astrophysik Potsdam (AIP),  
Max-Planck-Institut f\"ur Astronomie (MPIA Heidelberg), 
Max-Planck-Institut f\"ur Astrophysik (MPA Garching), 
Max-Planck-Institut f\"ur Extraterrestrische Physik (MPE), 
National Astronomical Observatories of China, New Mexico State University, 
New York University, University of Notre Dame, 
Observat\'ario Nacional / MCTI, The Ohio State University, 
Pennsylvania State University, Shanghai Astronomical Observatory, 
United Kingdom Participation Group,
Universidad Nacional Aut\'onoma de M\'exico, University of Arizona, 
University of Colorado Boulder, University of Oxford, University of Portsmouth, 
University of Utah, University of Virginia, University of Washington, University of Wisconsin, 
Vanderbilt University, and Yale University.

\end{document}